\documentclass{article}
\usepackage{epsfig}
\begin{document}

\begin{center}
\Large{\textbf{Realistic calculations of correlations and final state
      interaction effects in the $A(e,e'p)X$ process off complex nuclei}
      \footnote{\small Presented by M. Alvioli at the 
      \textit{2nd International Conference on Nuclear and Particle Physics
      with CEBAF at  JLab}, to appear in the Proceedings.}
      }
\end{center}

\begin{center}\Large{
 M. Alvioli$^*$, C. Ciofi degli Atti$^*$, H.
      Morita$^\dagger$}
\end{center}

\begin{center}
$*$ Department of Physics, University of Perugia, and Istituto
      Nazionale di Fisica Nucleare, Sezione di Perugia, Via A. Pascoli,
      I--06100 Perugia, Italy\\
$\dagger$ Sapporo Gakuin University, Bunkyo-dai 11, Ebetsu 069,
      Hokkaido, Japan
\end{center}

\noindent
A linked cluster expansion for the calculation of ground state observables 
of complex nuclei with realistic interactions has been used to calculate
the ground state energy, density and momentum distribution of $^{16}O$
and $^{40}Ca$. Using the same cluster expansion and the wave function and
correlation parameters obtained from the energy calculation, we have evaluated
the semi inclusive reaction $A(e,e'p)X$ taking final state interaction (FSI)
into account by  a Glauber type approach; the comparison between the distorted
and undistorted momentum distributions provides an estimate of the transparency
of the nuclear medium to the propagation of the hit proton. The effect of color
transparency is also included by considering  the Finite Formation Time (FFT)
that the hit hadron needs to reach its asymptotic  physical state.
\section{Introduction}
The exclusive, $A(e,e'p)B$, and semi-inclusive,  $A(e,e'p)X$,
electro-disintegration of nuclei  represent a powerful tool to investigate 
various aspects of nuclear structure (e.g. single-particle motion and mean field
effects, nuclear correlations),  as well as QCD motivated effects (e.g. color
transparency). The accuracy of recent and forthcoming experimental data
require realistic  theoretical calculations to be performed, based as much
as possible on a description of the nucleus stemming from \textit{first 
principle} calculations, which means that the nuclear wave functions which
appear in the calculation of various matrix elements,  used either to predict
or interpret the experimental data, should  in principle result from many body
calculations and realistic interactions.
The problem has been solved in the case of few body system, for which
realistic wave functions are currently being used in the interpretation
of electro-disintegration processes, but this problem still needs a 
solution in case of complex nuclei.  
As a matter of fact, calculations of ground state observables for complex
nuclei represent still a hard task, and even in those cases when  approximate
many body calculations for the ground state energy can be performed,
the structure of the wave function
is so complicated that its use for calculations of matrix elements of operators,
different from the potential and kinetic energies ones, is very involved. 
For such a reason, a simpler, but still realistic, method which would allow one
to calculate various kinds of matrix elements with nuclear wave functions which
correctly incorporate the most relevant features of a realistic wave function,
in particular  its correlation structure resulting from the main features of
modern two-nucleon interactions, would be extremely useful.
Cluster expansion techniques, when the expectation value of different operators
can be calculated  to a certain order,
may represent a valid and practicable alternative to the full complex "exact"
solution  of the many body problem. It is the aim of the present paper to
illustrate a cluster expansion approach to the calculation of ground state
energy properties (energy, density and momentum distribution) and various
types of electro-disintegration processes.
\section{Cluster expansion and the nuclear wave function}
In our linked-cluster expansion approach, the expectation value of a
certain operator $\hat{\mathcal{O}}$ 
\begin{equation}
\label{omedio1}
\langle\hat{\mathcal{O}}\rangle\,=\,\frac{\langle\Psi_A|\,\hat{\mathcal{O}}
\,|\Psi_A\rangle}{\langle\Psi_A | \Psi_A\rangle}
\end{equation} 
is evaluated with correlated wave functions of the following "classical" form
\begin{equation}
\label{psi1}
\Psi_A\,=\,\hat{F}({\bf r}_1,...,{\bf r}_A)\,\Phi_A({\bf r}_1,...,{\bf r}_A)\,,
\end{equation} 
where $\Phi_A$ is a mean field (Slater determinant) wave function, and  $\hat{F}$
a symmetrized (by the symmetrization operator $\hat{S}$) correlation
operator which generates $\textit{correlations}$ into the mean field wave function;
it  has the following general form
\begin{equation}
\label{corre1}
\hat{F}\,=\,{{\hat{S}}}\prod^A_{i<j}\hat{f}(r_{ij}) \,\,
\end{equation}
with
\begin{equation}
\hat{f}(r_{ij})=\sum_p\,{\hat
f}^{(p)}(r_{ij})\;\;\hspace{2cm}{\hat f}^{(p)}(r_{ij})=
      f^{(p)}(r_{ij})\,\hat{O}^{(p)}_{ij}\,
\label{effecorr}
\end{equation}
where the operators $\hat{O}^{(p)}$ are the same which appear in the 
two-nucleon interaction, having the form ( e.g. in case of  a $V8$-type interaction) 
\begin{equation}
\label{operator}
{\hat{O}}^{p=1,8}_{ij}=\left[1,\, {\bf \sigma}_i
      \cdot{{\bf \sigma}}_j,\,S_{ij},\,({\bf L} \cdot {\bf S})_{ij}\right]\otimes
      \left[1,\,{\bf \tau}_i\cdot\ {\bf \tau}_j \right]\,
\end{equation}
The central parts $f^{(p)}(r_{ij})$'s of the correlation function
${\hat f}^{(p)}$, reflect the radial behaviour of the various components
and their actual form is determined either by the minimization of the ground
state energy, or by other criteria. 

\begin{figure}[!h]
\centerline{
      \epsfysize=5.9cm\rotatebox{-90}{\epsfbox{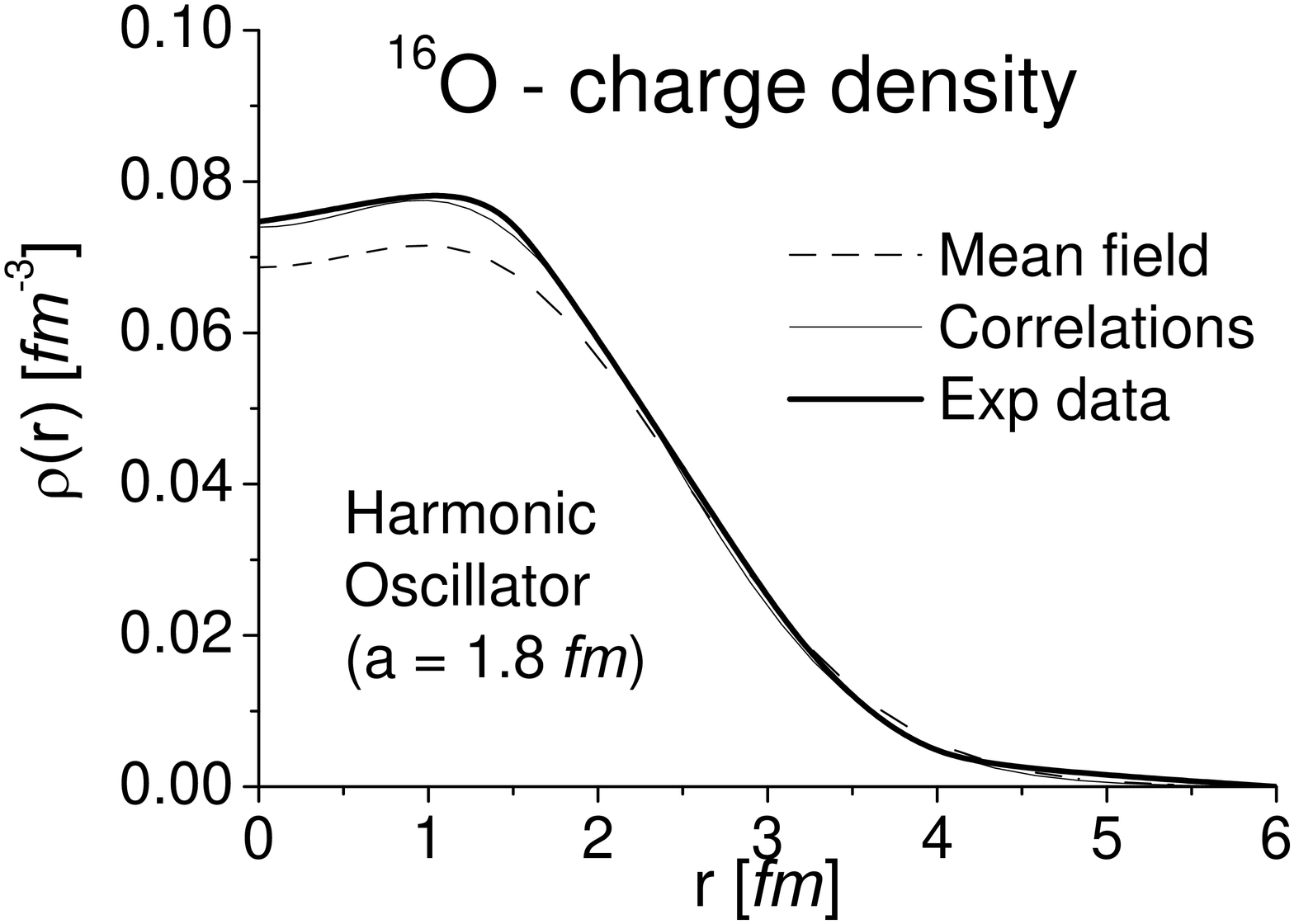}}
      \hspace{1mm}
      \epsfysize=6cm\rotatebox{-90}{\epsfbox{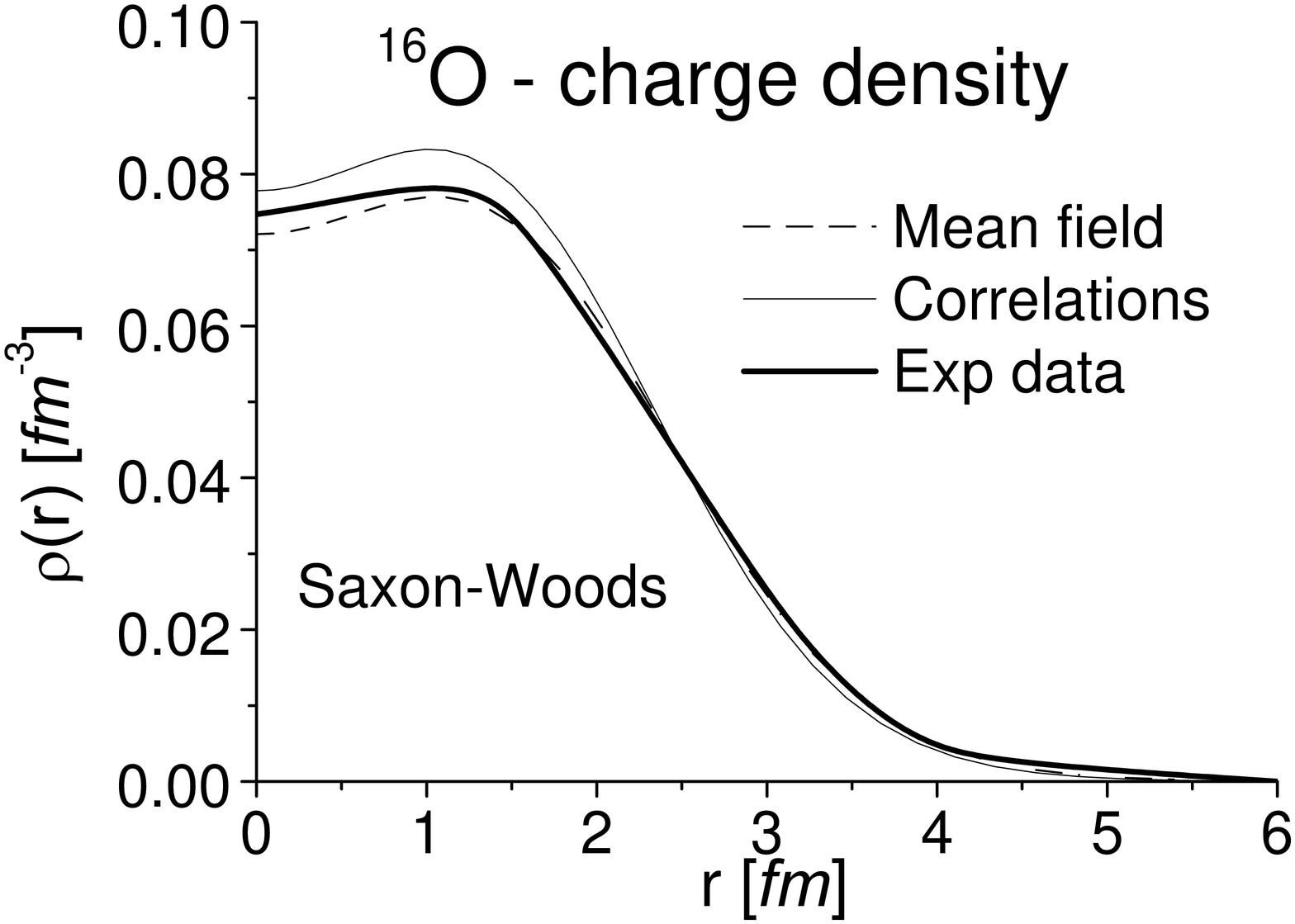}}
      }
\caption{The  charge densities of $^{16}O$ calculated using the cluster expansion
(\ref{eta1}) with Harmonic Oscillator (HO) (left panel)
and Saxon-Woods (SW) (right panel) single-particle wave functions ({\it spwf}).
\textit{Dashed line}: mean-field wave functions used in the cluster expansion
calculations; \textit{full line}: results of the cluster expansion;
\textit{thick full line}: experimental data \cite{dev01}.
The HO and the SW spwf's parameters corresponding to the full line have been
chosen such as to provide a good description of the density without appreciably
changing  the minimum value of the ground state energy. The normalization of the
density is $4\pi\int \rho(r)\,r^2\,dr = Z$, $Z$ being the number of protons.}
\label{density-opt}
\end{figure}

\begin{figure}[!ht]
\centerline{
      \epsfysize=7cm\rotatebox{-90}{\epsfbox{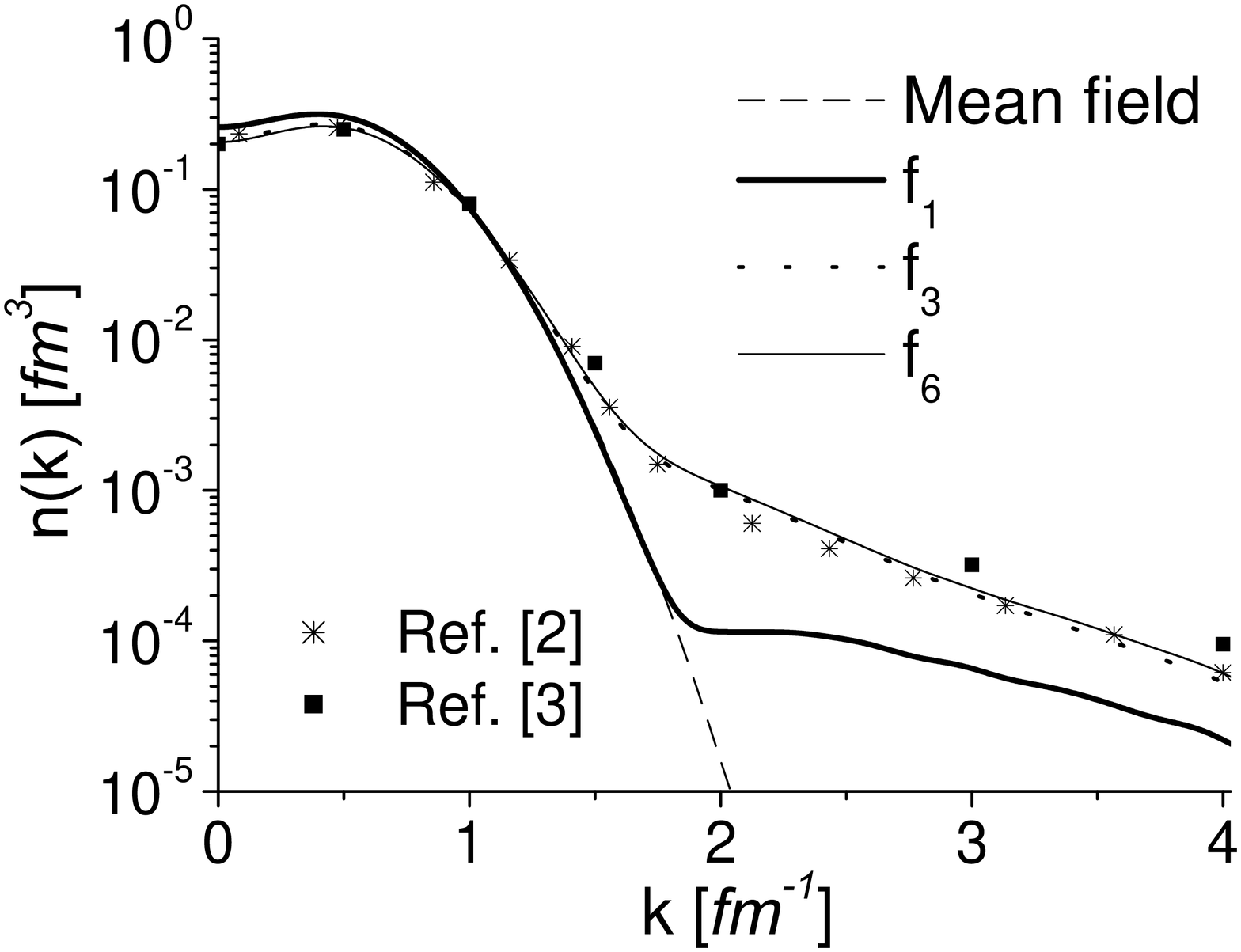}}
      }
\centerline{
      \epsfysize=7cm\rotatebox{-90}{\epsfbox{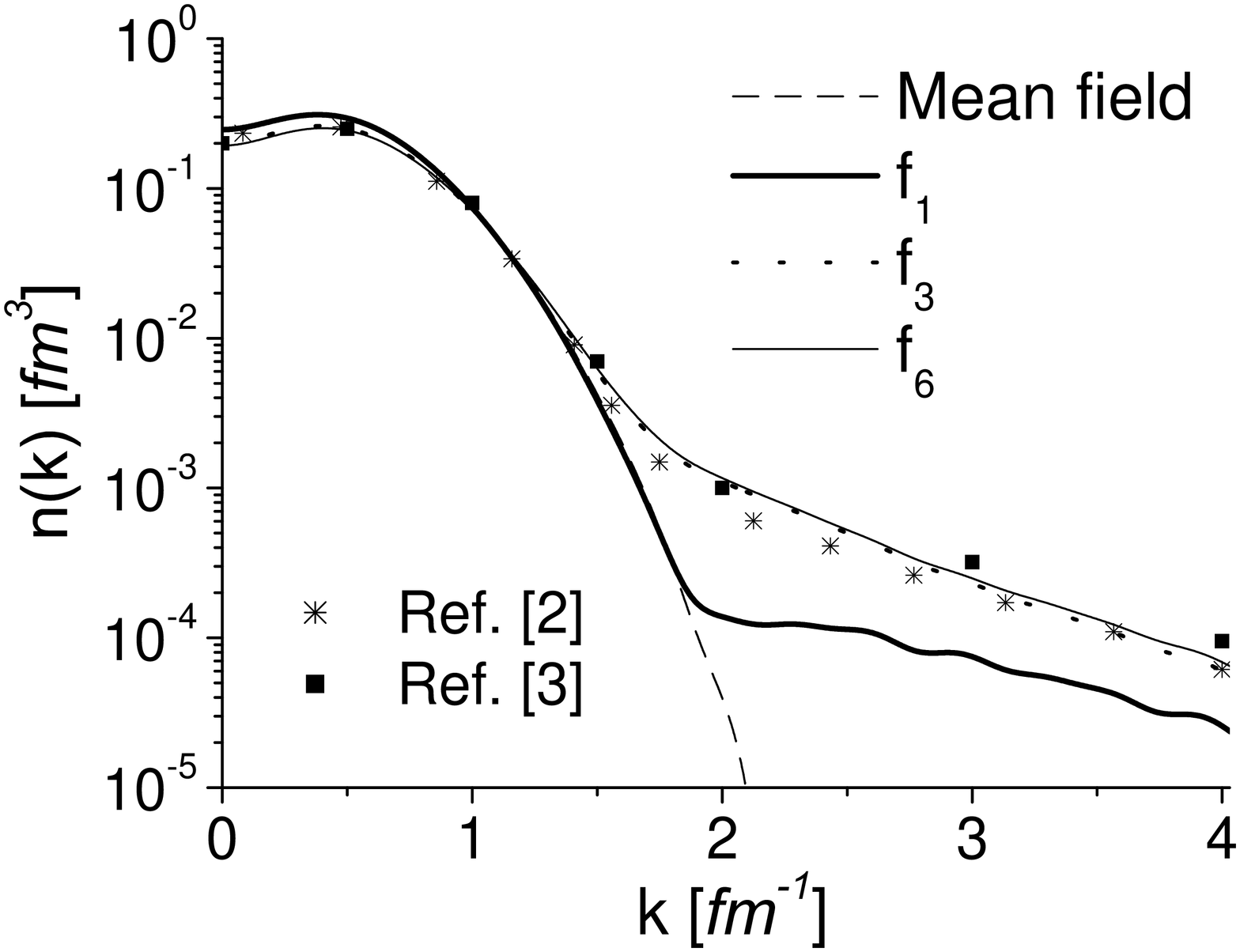}}
      }
\caption{The momentum distributions of  $^{16}O$  calculated with the cluster
expansion (\ref{eta1}) and Harmonic Oscillator (HO) (upper panel) and Saxon-Woods
(SW) (lower panel) \textit{spwf}'s as in Fig.1. \textit{Dashed line}: mean field
approximation; \textit{thick full line}: $f_1$ \textit{approximation} (only central
correlations: $(\hat{f}={\hat f}^{(1)}$); \textit{dotted line}: $f_3$
\textit{approximation} ($\hat{f}=\hat{f}^{(1)}+\hat{f}^{(4)} +\hat{f}^{(6)}$);
\textit{full line}: $f_6$ \textit{approximation}
($\hat{f}=\sum_{p=1}^6\,\hat{f}^{(p)}$).
The asterisks and the full squares are the results from Ref. \cite{fab01} and
\cite{pan01}, respectively. The values of the kinetic energy calculated by Eq.
\ref{kin1} are as follows: $<T>$ = 230.28 $MeV$ (\textit{dashes}),
287.23 $MeV$ (\textit{thick full}), 435.94 $MeV$ (\textit{dots}),
458.29 $MeV$ (\textit{full}) for H.O. \textit{spwf}'s, and $<T>$ = 244.94 $MeV$ 
(\textit{dashes}) 306.98 $MeV$ (\textit{thick full}), 469.83 (\textit{full}), 
494.47 $MeV$ (\textit{dots}) for S.W. \textit{spwf}'s.
The normalization of $n(k)$ is $4 \pi \int n(k)\,k^2\,dk\,=\,1$.}
\label{momdisHO16-opt}
\end{figure}

The cluster expansion of Eq.\ref{omedio1} is carried out in terms of the
quantity $\hat{\eta}_{ij}={\hat{f}}^2_{ij}-1$,  whose integral plays the
role of a small expansion parameter; by expanding the numerator and the
denominator  the terms $\hat{\mathcal{O}}_n$  of the same order $n$ in
$\eta_{ij}$, are collected obtaining $\langle\hat{\mathcal{O}}\rangle=
\mathcal{O}_0+\mathcal{O}_1+\mathcal{O}_2+...$, with
\begin{eqnarray}
\label{eta1}
\mathcal{O}_0&=&\langle\hat{\mathcal{O}}\rangle\, \nonumber\hspace{2.73cm}
      \mathcal{O}_1\,=\,\langle\sum_{ij}
      \hat{\eta}_{ij}\,\hat{\mathcal{O}}\rangle\,
      -\,\mathcal{O}_0\,\langle\sum_{ij}\,
      \hat{\eta}_{ij}\rangle\,\nonumber\\
\nonumber\\
\mathcal{O}_2&=&\langle\sum_{ij<kl}\hat{\eta}_{ij}\,\hat{\eta}_{kl}\,
      \hat{\mathcal{O}}\rangle\,
      -\,\langle\sum_{ij}\hat{\eta}_{ij}\,\hat{\mathcal{O}}\rangle\,
      \langle\sum_{ij}\hat{\eta}_{ij}\rangle\,+\nonumber\\
& &\hspace{3cm}+\,\mathcal{O}_0\,\left(\langle\sum_{ij<kl}\hat{\eta}_{ij}\,
      \hat{\eta}_{kl}\,\rangle\,-\langle\sum_{ij}\hat{\eta}_{ij}\rangle^2
      \right)\,;
\end{eqnarray}
where $\langle [...] \rangle \equiv \langle \Phi_A \left|[...] \right|\Phi_A
\rangle$. From now on, our approach will consist in obtaining the parameters
characterizing the correlation functions and the mean-field single-particle
wave function which correspond to an acceptable value  of the ground state energy,
we will then use the obtained wave function $\Psi_A$ to calculate the transition
matrix elements entering in the theoretical description of electro-disintegration
processes using  the same cluster expansion employed to calculate the energy.
We have calculated the ground state energy of $^{16}O$ and $^{40}Ca$
using the Argonne $V8'$ \cite{pud01} potential and adopting, as in Ref.
\cite{fab01}, the so called $f_6$ approximation consisting in considering
only the first six components of Eq. \ref{operator}.
The expectation value of the many body non relativistic Hamiltonian of the
nucleus was  obtained by calculating the average values of the kinetic and
potential energies, i.e.
\begin{equation}
\label{kin1}
\langle\hat{T}\rangle\,=\,-\frac{{\hbar}^2}{2m}\,\int\,d{\bf k}\,k^2\,n({k})\,,
\end{equation}
where $n(k)$ is the nucleon momentum distribution 
($k \equiv |{\bf k}|$), 
\begin{equation}
\label{momdis1}
n(k)\,=\,\frac{1}{(2\pi)^3}\,\int\,d{\bf r}_1\,d{\bf r}_1'
\,e^{-i\,{\bf k}\cdot({\bf r}_1-{\bf r}_1')}\,\rho^{(1)}({\bf r}_1,{\bf r}_1'),
\end{equation}
and
\begin{equation}
\label{pot1}
\langle\hat{V}\rangle\,=\,\frac{1}{2}\sum_{i<j}\langle \hat {v}_{ij}\rangle
=\,\frac{A(A-1)}{2}\sum_p\;\int\;d{\bf r}_1 d
      {\bf r}_2\;v^{(p)}(r_{12})\rho^{(2)}_{(p)}({\bf r}_1,{\bf r}_2)\,.
\end{equation}
The calculations have been performed by cluster expanding the expectation 
value of the non diagonal one-body, $\hat{\rho}^{(1)}$, and diagonal two-body, 
$\hat{\rho}^{(2)}({\bf r}_1,{\bf r}_2)$ density matrix operators. 
The six correlation functions $f^{(p)}(r_{ij})$ have been borrowed from  
Ref. \cite{fab01}, whereas  Harmonic Oscillator (HO) and Saxon-Woods (SW)
spwf's have been used to describe the mean field. 
As in  Ref. \cite{fab01}, we found that the charge densities
corresponding to the minimum of the energy, appreciably disagree with the
corresponding experimental quantities, therefore, in view of the mild  dependence
of  the energy around the minimum upon   the mean-field parameters, following
Ref. \cite{fab01} we have changed the latter to obtain agreement between
theoretical and experimental charge densities.
The results for the charge densities and momentum distributions, which
are shown in Figs. \ref{density-opt} and \ref{momdisHO16-opt},  deserve the
following comments:
\begin{enumerate}
\item{the agreement between our cluster expansion and FHC/SOC result of Ref. \cite{fab01}
is very good};
\item{both approaches predict momentum distributions which do not appreciably
differ from the ones obtained in Ref. \cite{pan01}, where the Variational Monte Carlo method
and the
$AV18$ interaction have been used;}
\item{the high momentum part of $n(k)$ is almost  entirely exhausted
by \textit{non-central}, \textit{long-range} correlations, with the
\textit{central}, \textit{short-range},  Jastrow correlations under-predicting
the high momentum part of $n(k)$ by about one order of magnitude;}
\item{the dominant non-central correlations are the isospin,
$f_4=f^{(4)}(r_{ij}) {\bf \tau}_i\cdot{\bf \tau}_j$, and isospin-tensor,  
$f_6=f^{(6)}(r_{ij}) {\bf \tau}_i\cdot{\bf \tau}_j S_{ij}$, correlations};
\end{enumerate}
\section{The final state interaction in $A(e,e'p)X$ reactions off complex
      nuclei: Glauber approach} 

\begin{figure}[!ht]
\centerline{
      \epsfysize=5.9cm\rotatebox{-90}{\epsfbox{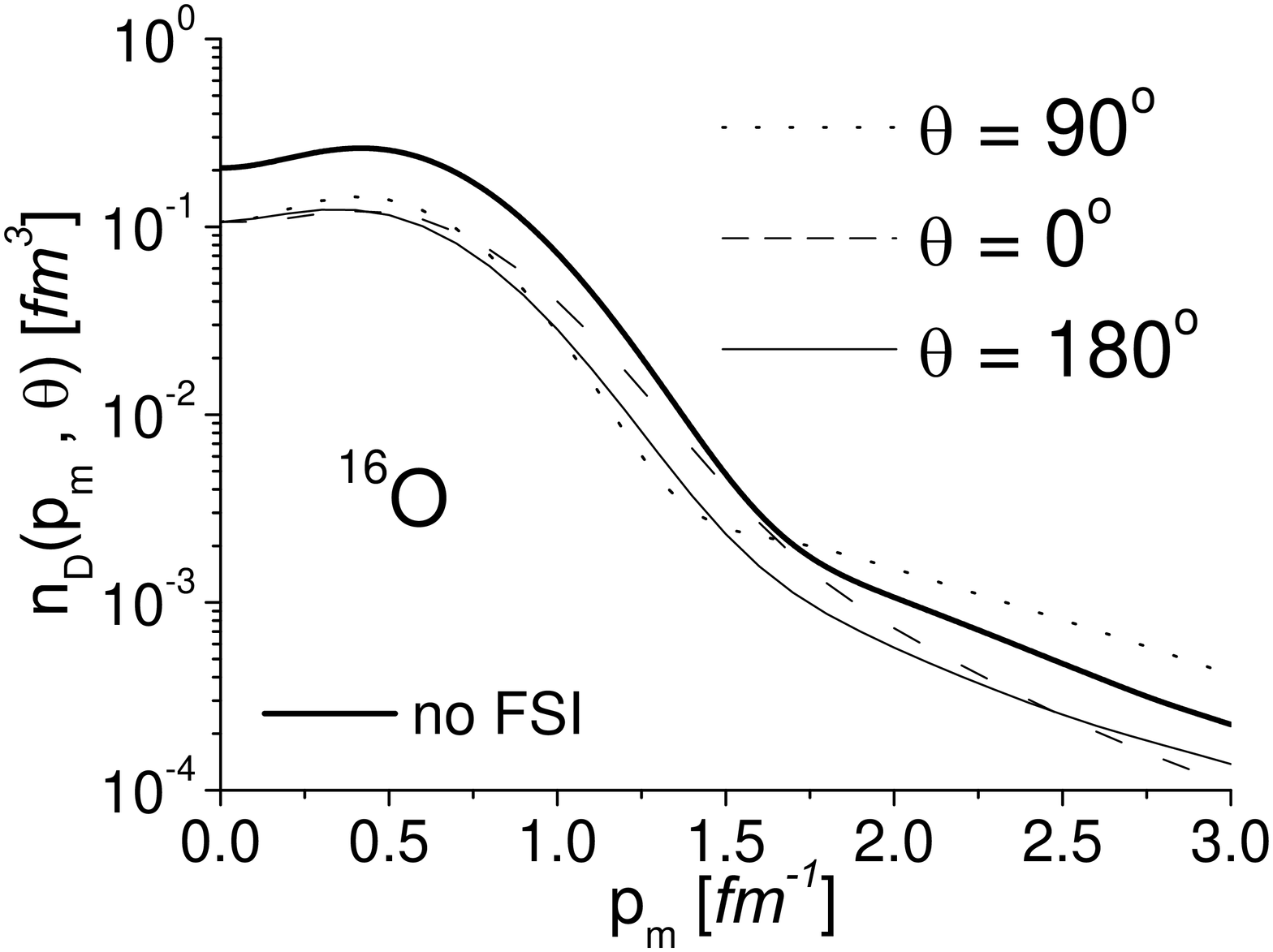}}
      \hspace{1mm}
      \epsfysize=5.9cm\rotatebox{-90}{\epsfbox{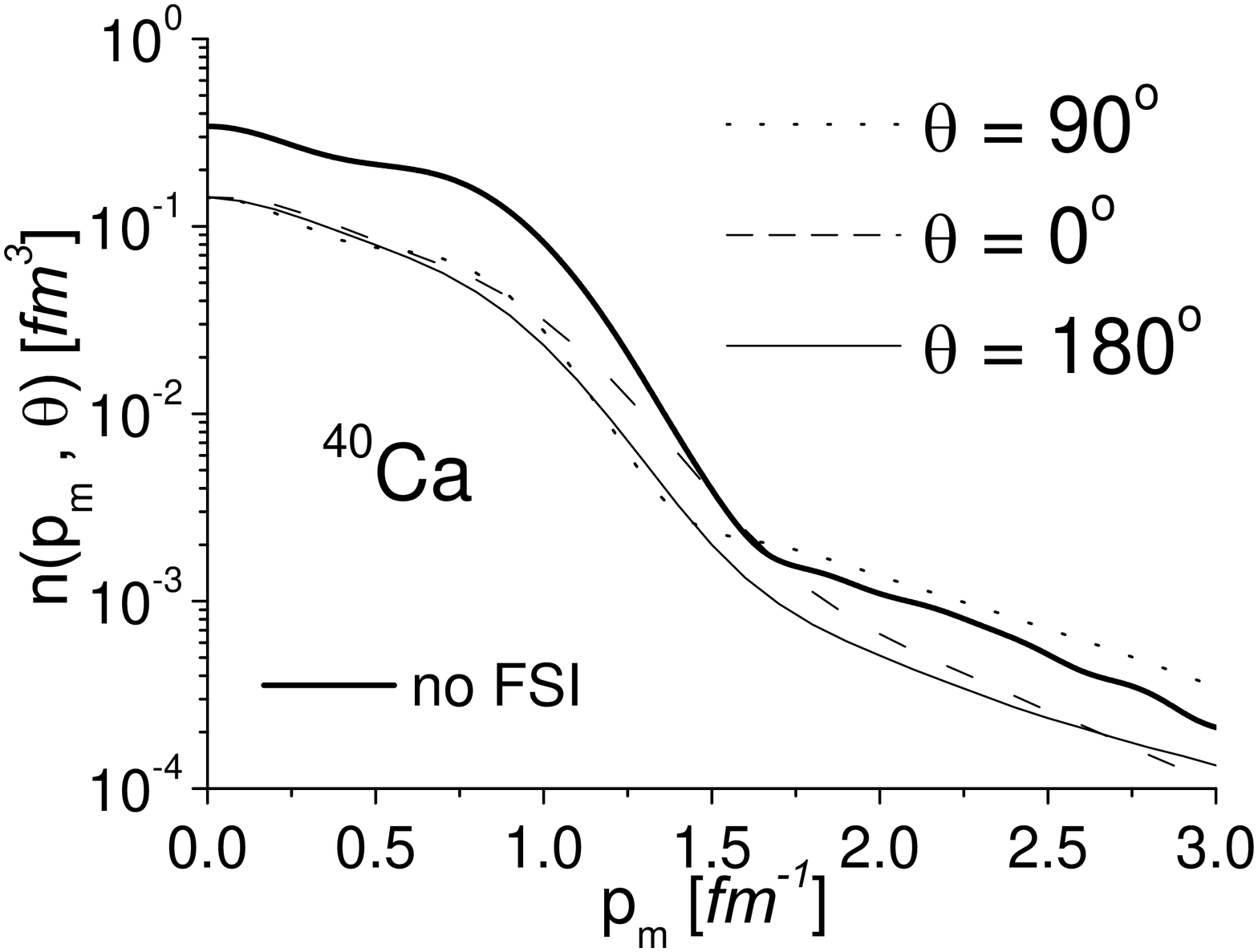}}
      }
\caption{The distorted momentum distribution, $n_D({\bf p}_m)=n_D(p_m, \theta)$ 
      ($\theta=\widehat{{\bf q}{\bf p}}_m$),  of $^{16}O$  and $^{40}Ca$,
      obtained  by Eq. \ref{nd} with correlated wave functions, Harmonic Oscillator
      ${\it spwf}'s$ and the Glauber $S$ matrix (\ref{SG}). The value of the
        integrated nuclear transparency (\ref{intnd}) for $^{16}O$ is $0.5$.}
      \label{momdisto01}
\end{figure}

Using the results obtained in the previous Section, we have calculated the
semi-inclusive $A(e,e'p)X$ process in which an electron with 4-momentum
$k_1\equiv\{{\bf k}_1,i\epsilon_1\}$, is scattered off a nucleus with 4-momentum
$P_A\equiv\{{\bf 0},iM_A\}$ to a state $k_2\equiv\{{\bf k}_2,i\epsilon_2\}$ 
and is detected in coincidence with a proton $p$ with 4-momentum 
$p\equiv\{{\bf p},iE_p\}$; the final $(A-1)$ nuclear system with 
4-momentum $P_X\equiv\{{\bf P}_X,iE_X\}$ is undetected. The cross section 
for the exclusive process $A(e,e'p)B$  can be written as follows
\begin{equation}
\frac{d\sigma}{dQ^2 d\nu d{\bf p}}=K\sigma_{ep}P_D(E_m,{\bf p}_m)
\label{sezione}
\end{equation}
where $K$ is a kinematical factor, $\sigma_{ep}$ the off-shell electron-nucleon 
cross section,  and $Q^2=|{\bf q}|^2-\nu^2$ the four momentum transfer. The 
quantity $P_D(E_m,{\bf p}_m)$ is the distorted nucleon spectral function which 
depends upon the observable \textit{missing momentum} ${\bf p}_m={\bf q}-{\bf p}$
(${\bf p}_m={\bf k}$ when the FSI is absent)
and \textit{missing energy} $E_m=\nu-T_p -T_{A-1}$. 
In the semi-inclusive $A(e,e'p)X$ process,   the cross section (\ref{sezione}) is
integrated over the missing energy $E_m$, at fixed value of ${\bf p}_m$ and 
becomes directly proportional to the \textit{distorted} 
momentum distribution
\begin{equation}
n_D({\bf p}_m)={(2 \pi)^{-3}} \int e^{i {\bf p}_m({\bf r}_1 -{\bf r}_1')}
      \rho_D ({\bf r}_1,{\bf r}_1') d{\bf r}_1 d{\bf r}_1'
\label{nd}
\end{equation}
where 
\begin{eqnarray}
\rho_D ({\bf r}_1,{\bf r}_1')= \frac {\langle\Psi_A\,S^{\dagger}
\,\hat{O}({\bf r}_1,{\bf r}_1')\,S'\,{\Psi_A}'\rangle}{\langle\Psi_A\Psi_A\rangle}
\label{rodi}
\end{eqnarray}
is the  distorted one-body mixed density matrix, 
$S$ the S-matrix describing FSI,  and  the primed quantities have
to  be evaluated at ${\bf r}_i'$ with $i=1, ...,A$. The integral of 
$n_D({\bf p}_m)$ gives the nuclear transparency $T$ 
\begin{equation}
T = \frac{\int n_D({\bf p}_m) d{\bf p}_m}{\int n(k) d{\bf k}} = 
      \int \rho_D ({\bf r})d{\bf r} = 1+ \Delta T
\label{intnd}
\end{equation}
where $\rho_D({\bf r})=\rho_D ({\bf r}_1={\bf r}'_1\equiv {\bf r})$
and  $\Delta T$ originates from the FSI.
In Ref. \cite{cio01} Eq. \ref{nd} has been evaluated using a Glauber
representation for the scattering matrix $S$, \textit{viz}
\begin{equation}
S \rightarrow S_G({\bf r}_1\dots{\bf r}_A)=\prod_{j=2}^AG({\bf r}_1,{\bf r}_j)
      \equiv \prod_{j=2}^A\bigl[1-\theta(z_j-z_1)\Gamma({\bf b}_1-{\bf b}_j)\bigr]
\label{SG}
\end{equation}
where ${\bf b}_j$ and $z_j$ are the transverse and the longitudinal components 
of the nucleon coordinate ${\bf r}_j\equiv({\bf  b}_j,z_j)$, ${\mit\Gamma}({\bf b})$
the Glauber profile function for elastic proton nucleon scattering, and the 
function $\theta(z_j-z_1)$ takes care of the fact that the struck proton
``1''  propagates along a straight-path trajectory so that it interacts with
nucleon  ``$j$'' only if $z_j>z_1$. The same cluster expansion described in
Section II has been used taking Glauber rescattering exactly into account at
the given order $n$, and using the approximation $|\Psi_{A-3}|^2 =
\prod_3^A \rho(i)$.
Using the mean-field and correlation parameters obtained from the
energy calculation, we have obtained the  \textit{distorted} nucleon momentum
distributions $n_D({\bf p}_m)=n_D(p_m, \theta)$, where $\theta$ is the angle
 between ${\bf q}$ and ${\bf p}_m$; the results for     $^{16}O$ and $^{40}Ca$
are presented in Fig. \ref{momdisto01}.
\section{Finite formation time  effects}
Recently \cite{bra01} the effects of color transparency in
quasi-elastic lepton scattering off nuclei, have been introduced
by explicitly considering the finite formation time (FFT) that the
hit hadron needs to evolve to its asymptotic physical state.
It has been shown that at the values of the Bjorken scaling variable 
$x=Q^2/2m\nu \simeq 1$, FFT effects can be treated in a simple way,
i.e. by replacing the Glauber operator (Eq. \ref{SG})
with
\begin{figure}[!h]
\centerline{
      \epsfysize=5.9cm\rotatebox{-90}{\epsfbox{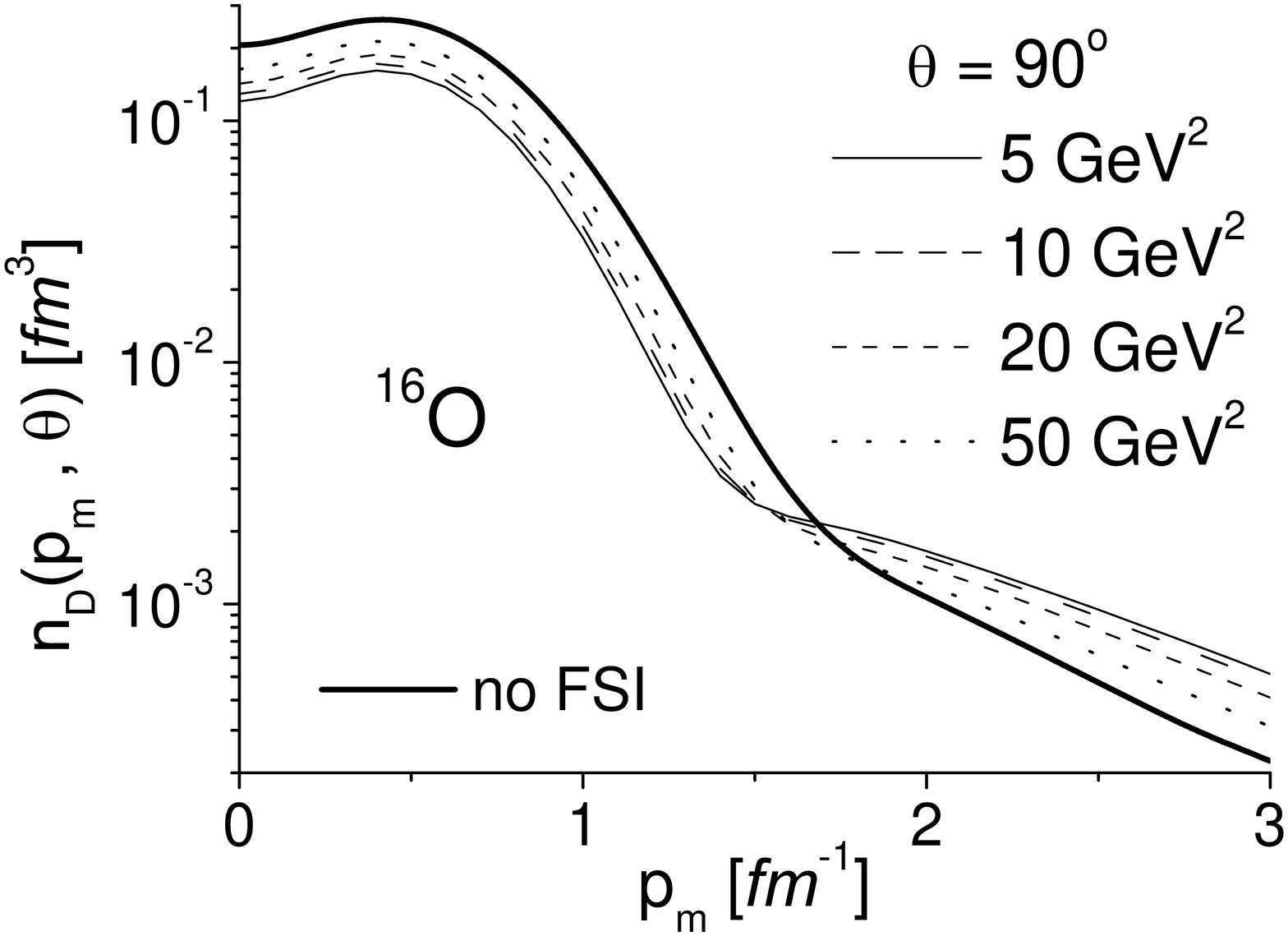}}
      \hspace{1mm}
      \epsfysize=5.6cm\rotatebox{-90}{\epsfbox{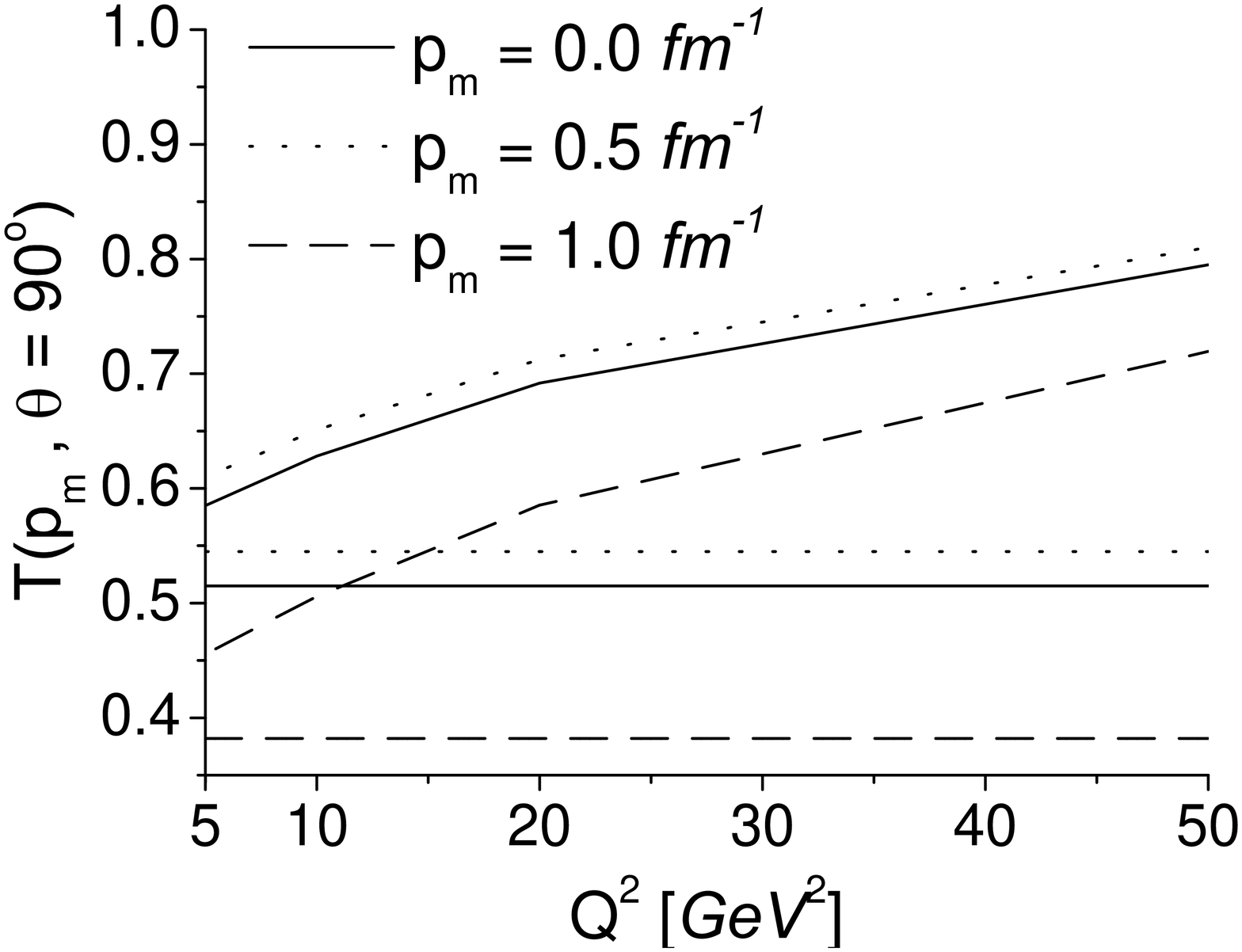}}
      }
\caption{\textit{Left panel}: the  $Q^2$- and $p_m$-dependence of  FFT effects
 on the distorted momentum distribution of $^{16}O$ calculated by Eqs. \ref{nd}
 and \ref{SG} at $\theta =90^o$. \textit{Right panel}: the ratio of the distorted
 to the undistorted momentum distributions, \textit{i.e.} the non-integrated nuclear
 transparency for various values of $p_m$ and $\theta=90^o$. The horizontal
 lines represent the $Q^2$-independent Glauber results.}
      \label{transpa02}
\end{figure}
\begin{equation}
\label{smatrix}
S_{FFT}({\bf r}_1,...,{\bf r}_A)=\,
      \,\prod^A_{j=2}\Big(1-J(z_1-z_j)\Gamma({\bf b}_1-{\bf b}_j)\Big)\,,
\end{equation}
where
\begin{equation}
J(z)\,=\,\theta(z)\,e^{-\,z\,\frac{x\,m\,M^2}{Q^2}}\,
\end{equation}
$m$ being  the nucleon mass  and $M^2=m^{*\,2}-m^2$ is a parameter 
describing the average excitation energy of the ejectile. It can be seen that 
 at sufficiently high values of $Q^2$,  $J \rightarrow  1$ and the FSI vanishes.
 The effects of FFT on the distorted momentum distribution is shown in Fig. \ref{transpa02}.

\section{Summary and conclusions}

We have obtained fully correlated wave functions by calculating the average value
of the nuclear Hamiltonian by means of a linked cluster expansion and using realistic
two-nucleon interactions. The wave functions have been used to obtain the ground
state density and momentum distribution. By introducing FSI effect by a Glauber-type 
approach, the distorted momentum distributions appearing in the semi-inclusive
$A(e,e'p)X$ processes have been calculated. By such a procedure, a consistent
treatment of initial-state correlations and final-state interactions has been
achieved. Color transparency effects have also been investigated by the Finite
Formation Time approach. Comparison with available experimental data are in
progress and will be reported elsewhere.\\

This work is part of the research activity of M. A. performed, under the supervision
of C. d. A., for the fulfillment of the PhD title. H. M. thanks the Department
of Physics, University of Perugia, and INFN, Sezione di Perugia, for hospitality
and support. Partial support by the Italian Ministero dell'Istruzione, 
Universit\'a e Ricerca (MIUR), through the funds COFIN01, is acknowledged.
We are indebted to A. Fabrocini for providing the correlation functions
obtained in Ref. \cite{fab01}.

\newpage
\bibliography{apssamp}

\begin{thebibliography}{3}
\bibitem[1]{pud01} B.S. Pudliner, V.R. Pandharipande, J. Carlson,
      S.C. Pieper and R.B. Wiringa,
      \textit{Phys. Rev.} \textbf{C56}, 1720 (1997);   
\bibitem[2]{fab01} A. Fabrocini, G. Co', 
      \textit{Phys. Rev.} \textbf{C61}, 044302 (2000)
      and Private Communication;
\bibitem[3]{pan01} S.C. Pieper, R.B. Wiringa and V.R. Pandharipande,
      \textit{Phys. Rev.} \textbf{C46}, 1741 (2000);   
\bibitem[4]{dev01} H. de Vries, C.W. de Jager and C. de Vries,
            \textit{Atom. Data Nucl. Data Tabl.} \textbf{36}, 495 (1987)
\bibitem[5]{cio01} C.Ciofi degli Atti, D. Treleani
      \textit{Phys. Rev.} \textbf{C60}, 024602 (1999);
\bibitem[6]{bra01} M.A. Braun, C.Ciofi degli Atti, D. Treleani,
      \textit{Phys. Rev.} \textbf{C62}, 034606 (2000);
\bibitem[7]{mor01} H. Morita, M.A. Braun, C.Ciofi degli Atti, D. Treleani
      \textit{Nucl. Phys.} \textbf{A699}, 328c (2002);
\bibitem[8]{alv01} M. Alvioli, C.Ciofi degli Atti, H. Morita
      \textit{(in preparation)}
\end{thebibliography}
\end{document}